\documentclass[conference]{IEEEtran}
\IEEEoverridecommandlockouts
\usepackage{bm}      % for \bm macro
\usepackage{amsmath,amssymb,amsfonts}
\usepackage{array}
\usepackage{algorithmic}
\usepackage{booktabs}
\usepackage{glossaries}
\usepackage{glossary-longbooktabs}
\usepackage{pgfplots}
\pgfplotsset{compat=1.18}
\usepackage{multirow}
\usepackage{tablefootnote}
\usepackage{graphicx}
\usepackage{textcomp}
\usepackage{xcolor}
\usepackage{lipsum}
\usepackage{xparse}
\usepackage{nomencl}
\usepackage{etoolbox}

\makeatletter
\let\MYcaption\@makecaption
\makeatother
\usepackage{subcaption}
\makeatletter
\let\@makecaption\MYcaption
\makeatother
\makenomenclature

\def\BibTeX{{\rm B\kern-.05em{\sc i\kern-.025em b}\kern-.08em
    T\kern-.1667em\lower.7ex\hbox{E}\kern-.125emX}}

\renewcommand\nomgroup[1]{%
  \item[\bfseries
  \ifstrequal{#1}{A}{Sets and Indices}{%
  \ifstrequal{#1}{B}{Decision Variables}{%
  \ifstrequal{#1}{C}{Auxiliary Variables}{%
  \ifstrequal{#1}{D}{Random Variables}{%
  \ifstrequal{#1}{E}{Parameters}{%
  }}}}}%
  ]
}

\begin{document}

%%%%%% SYMBOLS %%%%%%

\newcommand{\indexscenario}{\ensuremath{\omega}}

\newcommand{\setscenario}{\ensuremath{\Omega}}

\newcommand{\indextime}{\ensuremath{t}}

\newcommand{\settime}{\ensuremath{T}}

\newcommand{\indexsegment}{\ensuremath{i}}
\newcommand{\nsegments}{\ensuremath{N}}

\newcommand{\price}{\ensuremath{\lambda}}
\newcommand{\power}{\ensuremath{p}}

% Random variables
\NewDocumentCommand{\rvprice}{m O{\indexscenario} O{\indextime}}{\
  \ensuremath{\price_{#1}^{#3 #2}}}

\NewDocumentCommand{\rvwind}{O{\indexscenario} O{\indextime}}{
  \ensuremath{\power_{max}^{#2 #1}}
}

% Decision variables
\NewDocumentCommand{\varsegmentactive}{O{\indexsegment} O{\indexscenario} O{\indextime} }{
  \ensuremath{u^{#3 #2 #1}}
}

\NewDocumentCommand{\varsegmentpower}{
  O{\indexsegment} O{\indextime} }{
  \ensuremath{\power^{#2 #1}_{offer}}
}

\NewDocumentCommand{\varsegmentprice}{
  O{\indexsegment} O{\indextime} }{
  \ensuremath{\price^{#2 #1}_{offer}}
}

% Auxiliary Variables

\NewDocumentCommand{\auxshortfall}{O{\indexscenario} O{\indextime} }{
  \ensuremath{\Delta^{#2 #1}}
}
\NewDocumentCommand{\auxcleared}{O{\indexscenario} O{\indextime} }{
  \ensuremath{\power_{CL}^{#2 #1}}
}

% Parameters
\newcommand{\parambeta}{\ensuremath{\beta}}

\title{A Mixed Integer Quadratic Program for Valuing the Impact of Price and
Forecast Uncertainty for Wind Generators

\thanks{The information, data, or work
presented herein was supported by the National Science Foundation Graduate
Research Fellowship under Grant No. 2141064 as well as in part by the Advanced
Research Projects Agency-Energy (ARPA-E), U.S. Department of Energy under Award
Number DE-AR0001277. The views and opinions of authors expressed herein do not
necessarily state or reflect those of the United States Government or any
agency thereof.} }

\author{\IEEEauthorblockN{Daniel Shen and Marija Ilic}
\IEEEauthorblockA{\textit{Department of Electrical Engineering and Computer Science} \\
\textit{Massachusetts Institute of Technology}\\
Cambridge, USA \\
\texttt{oski@mit.edu} and \texttt{ilic@mit.edu}}
}

\maketitle

\begin{abstract}
  Owners of wind power plants are exposed to financial risk in wholesale
  electricity markets due to the uncertain nature of wind forecasts and price
  volatility. In the event of a wind shortfall, the plant may have to
  repurchase power at a higher price in the real-time market. However, reducing
  the power offered in the day-ahead market may also be interpreted by
  regulators as physical withholding. We formulate and solve a mixed-integer
  quadratic program (MIQP) that prices the uncertain portion of a wind
  generator’s forecast to hedge against uncertainties and which addresses
  concerns around withholding. We exploit the structure of the MIQP inputs to
  introduce additional constraints to improve computation time. Additionally,
  we provide a qualitative approach for generators and regulators to interpret
  the results of the MIQP. Finally, we simulate a real-world application for a
  wind farm in New York using past wind forecasts and NYISO prices.
\end{abstract}


%%%%% NOMENCLATURE %%%%%
% Nomenclature sections:
% Variable and Parameter Indexing (A)
% Decision Variables (B)
% Auxiliary Variables (C)
% Random Variables (D)
% Parameters (E)

\nomenclature[A, 01]{\setscenario}{Set of price and wind forecasts.}
\nomenclature[A, 02]{\indexscenario}{Index over forecast scenarios.}
\nomenclature[A, 03]{\indextime}{Time indexer.}
\nomenclature[A, 04]{\indexsegment}{Offer curve segment indexer.}
\nomenclature[A, 05]{\nsegments}{Total number of offer curve segments.}

\nomenclature[B, 01]{$\varsegmentpower$}{
  Power offered in segment \indexsegment.
}
\nomenclature[B, 02]{$\varsegmentprice$}{
  Price of the power offered in segment \indexsegment.
}

\nomenclature[C, 01]{\varsegmentactive}{
  Indicator if segment $\indexsegment$ clears the market.
}
\nomenclature[C, 01]{\auxcleared}{
  Power cleared in day-ahead under scenario \indexscenario.
}
\nomenclature[C, 01]{\auxshortfall}{
  Shortfall between the day-ahead clearing amount and the real-time forecast
  realization; negative if there is a repurchase.
}

\nomenclature[D, 01]{\rvprice{DA},\rvprice{RT}}{
  Day-ahead and real-time prices in scenario \indexscenario.
}
\nomenclature[D, 02]{\rvwind}{
  Max. dispatchable wind power in scenario \indexscenario.
}
\nomenclature[E, 01]{\parambeta}{Risk factor for conditional value at risk.}
\nomenclature[E, 02]{$M$}{
  Arbitrarily large constant for indicator constraints.
}
\nomenclature[E, 03]{$\epsilon$}{
  Arbitrarily small constant for indicator constraints.
}

\printnomenclature[1.8cm]

\section{Introduction}

Significant amounts of variable renewable energy (VRE) must be installed on the
grid to achieve decarbonization targets. Such resources incur no fuel costs to
produce electricity, but their generation output is uncertain and the asset
owners face financial risk from underdelivery of power in the real-time market.
It is typically assumed in restructured electricity markets that generators
should offer power at a price that reflects their variable costs. While the
fuel portion of variable cost for VRE resources is zero, VRE generators do
incur nonzero costs in potential shortfall penalties as the volume of their
power sold on the day-ahead market increases. In the scenario where the
real-time electricity price is higher than the day-ahead electricity price, a
wind generator which always offers and clears at its full rated capacity will
incur costs from consistently repurchasing its real-time shortfall.

For work done previously on price-taker behavior offer/bid curve creation for
producers and consumers, we refer to \cite{curves-wind-bids} and
\cite{curves-demand} \cite{curves-price-taker-norway}, respectively. Our work
builds upon the approach described in \cite{yin-risk-based-dispatch}, but for
VRE deviations instead of dispatch deviations by conventional generators.
This work also extends previous work on wind generator offer creation in
\cite{shen-acc} to a piecewise linear context. We refer to
\cite{wind-power-offer-strategy-shin} for a detailed treatment of strategic
wind power offering with conditional value at risk.

This work extends some of the aforementioned work by combining the concept of
breaking the problem down into discrete block price-quantity segments, as is
required in nearly all restructured markets, and the concept of optimizing
those segments under the conditional value at risk (CVaR) risk measure. We
consider the problem of a price-taking wind generator pricing its variable
costs associated with uncertainty in a day-ahead offer curve. We propose a
mixed-integer quadratic program (MIQP) that considers CVaR to reflect the
variable costs of shortfall penalties. The MIQP takes into account the
distribution of day-ahead prices, real-time prices, and wind forecasts to
create an offer curve of energy blocks at specific marginal prices.

We wish to note the following contributions of this paper:
\begin{itemize}
  \item Formulating a MIQP program which optimizes both the price and quantity
  dimensions of the generator offer curve to reflect the variable costs of
  uncertainty.
  \item Reducing the number of unconstrained integer decision variables in the
  MIQP by exploiting structure in the optimization problem to improve
  computational speed.
  \item Presenting an approach for qualitatively interpreting the offer curve
  results by vizualizing the distributions of the ``active'' price and wind
  scenarios. Such a method could prove useful for generation operators to
  justify their nonzero marginal cost offers to other stakeholders or market
  administrators.
\end{itemize}

\section{Methodology}

\subsection{Problem Setup \& Background}
\label{sec:setup}
In restructured electricity markets, generators submit offer curves of blocks
of power offered at discrete marginal prices. We make the following
considerations for a wind generator:

\begin{itemize}
  \item The agent takes the day-ahead and real-time price scenarios as
    predicted (price-taker behavior) and also has a set of discrete forecast
    scenarios of its real-time wind production.
  \item All power quantities offered at a price below the day-ahead price
    clear the market and are sold at that day-ahead price.
  \item In the day-ahead optimization, the agent should not
    consider any profits from selling or repurchasing electricity in the real-time
    market to counteract claims that the generator is engaging in physical
    withholding \cite{market-manipulation} and could potentially be exercising
    market power.
  \item If the realized wind power is less in real-time than was contracted
    for dispatch in the day-ahead market, the generator must buyback the
    shortfall at the real-time price.
  \item The generator is downwards dispatchable to 0 MW from any realized wind
    scenario and such downwards dispatch incurs no operating cost.
\end{itemize}

\begin{figure}[!b]
  \begin{center}
      \scalebox{0.7}{\input{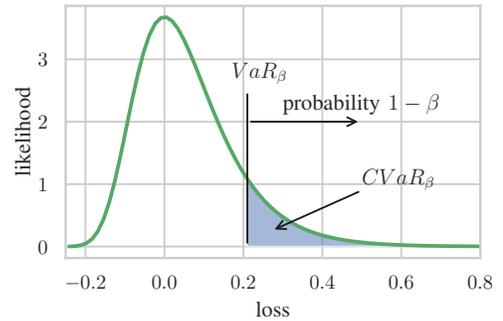}}
  \end{center}
  \vspace*{-5mm}
  \caption{Graphical representation of value at risk (VaR) and conditional
    value at risk (CVaR).}
  \label{fig:cvar-explain}
\end{figure}

Electricity prices are generally more volatile than that of other commodities,
and generators may have a preference to reduce the risk of incurring financial
losses from price fluctuations. We integrate a popular risk measure,
conditional value at risk (CVaR) \cite{cvar-paper}, into our objective to allow
for modeling of the agent's aversion to risk. In using CVaR, we only consider
the most undesired scenarios as determined by the choice of offer curve. Fig.
\ref{fig:cvar-explain} gives a graphical representation of CVaR. The CVaR
parameter $\parambeta$ represents the agent's aversion to risk; as $\parambeta$
increases up to 1, the agent's aversion to risk also increases. Using a
sample-average approximation approach with 100 scenarios, the CVaR is equal to
the $(1-\parambeta) * 100$ lowest- profit scenarios. For $\parambeta=0$, the
problem is equivalent to optimizing the expectation.

\subsection{Mixed-Integer Quadratic Program}
\label{sec:miqp}

Market operators require the offer curve of the generator to be a piecewise
linear function of total production costs (\$/h) vs. output (MW) for the market
clearing software. This is equivalent to a step function that specifies the
marginal cost (\$/MWh) as a function of output for $\nsegments$ steps. For the
remainder of this paper, the term ``offer'' will refer to the latter step
function.

\begin{samepage}

For a single period $\indextime$, the optimal offer based on considerations
in Section \ref{sec:setup} is found by solving the following MIQP:

\begin{subequations}
  \begin{equation}
    \max_{\bm{x}} \quad CVaR_{\parambeta}
    \left(
      \rvprice{DA}[] \: \auxcleared[] + \rvprice{RT}[] \auxshortfall[]
    \right) \label{eq:objective}
  \end{equation}
  \begin{align}
    \text{subject to}\quad
      & \auxcleared = \sum_{\indexsegment}^{\nsegments}
        \varsegmentactive \varsegmentpower \label{eq:cleared} \\
      & max(\auxcleared) \leq max(\rvwind) \label{eq:pmax} \\
    & \auxshortfall = min(0, \rvwind - \auxcleared) \label{eq:shortfall} \\
    & \varsegmentactive \leq 1 - \frac{\varsegmentprice - \rvprice{DA}}{M}
      \label{eq:active_binary_zero} \\
    & \varsegmentactive \geq \epsilon +
      \frac{\rvprice{DA} - \varsegmentprice}{M}
      \label{eq:active_binary_one}\\
    & \varsegmentactive \in \{ 0, 1 \} \label{eq:active_binary_binary} \\
    & \sum_i^{\nsegments} \varsegmentactive \leq
      \sum_i^{\nsegments} \varsegmentactive[\indextime][(\indexscenario + 1)]
      \label{eq:indicator-rowsum} \\
    &
     \varsegmentprice[\indexsegment] \leq \varsegmentprice[(\indexsegment + 1)] \label{eq:increasingprice} \\
    & \varsegmentactive[\indextime][\indexscenario][(\indexsegment + 1)] \leq
      \varsegmentactive \label{eq:indicator-decreasing}
  \end{align}
  \begin{equation*}
    \text{where} \quad \bm{x} =
      \varsegmentactive, \varsegmentprice, \varsegmentpower
  \end{equation*}
\end{subequations}

\end{samepage}

\begin{figure}[tb]
  \begin{center}
    \hspace*{-1cm}
    \includegraphics[width=0.9\columnwidth]{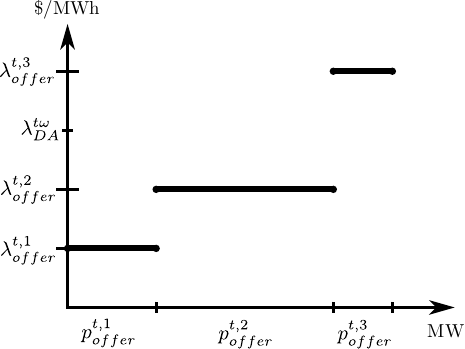}
  \caption{
    Representation of offer curve parameters. Observe that based on the
    indicated forecast price in scenario \indexscenario, only segments 1 and 2
    would clear and
    $[\varsegmentactive[1], \varsegmentactive[2], \varsegmentactive[3]] = [1, 1, 0]$
    }
    \label{fig:offer-curve}
  \end{center}
\end{figure}

Variables $\varsegmentprice$ and $\varsegmentpower$ specify the price and
quantity characteristics of each block in the offer curve
(Fig. \ref{fig:offer-curve}). $\varsegmentactive$ is 1 if a block
$\indexsegment$ clears in a specific scenario and 0 otherwise.

The objective is to maximize the revenue cleared in the day-ahead market
subject to the $CVaR_\parambeta$ risk measure. Each scenario $\indexscenario$
consists of a hourly day-ahead price sample $\rvprice{DA}$, real-time price
sample $\rvprice{RT}$, and a forecast for the maximum wind power $\rvwind$.

The constraints can be interpreted as follows:
\begin{itemize}
  \item (\ref{eq:cleared}) gives the quantity of power cleared in
  the day-ahead market under scenario $\indexscenario$.
  \item (\ref{eq:pmax}) restricts the generator from submitting a total
    quantity offer that exceeds the maximum forecast scenario.
  \item (\ref{eq:shortfall}) restricts the optimization from considering
  net profit obtained from withholding power for the real-time market; the
  variable $\auxshortfall$ is negative if the generator is buying back power to
  cover a shortfall.
  \item (\ref{eq:active_binary_zero}), (\ref{eq:active_binary_one}), and
  (\ref{eq:active_binary_binary}) reflect which of the offered blocks clear
  for a given scenario in the day-ahead market; is is assume that if the market
  price is equal to the offer price, the entire quantity associated with the
  block clears. For a graphical representation of this constraint, see
  Fig. \ref{fig:offer-curve}.
  \item (\ref{eq:indicator-rowsum}) follows from assuming the scenarios are
    presorted by increasing day-ahead prices.
  \item (\ref{eq:indicator-decreasing}) follows from
    (\ref{eq:increasingprice}), since lower-priced blocks will always clear
    if a higher-priced block clears as well.
\end{itemize}

Note that constraints (\ref{eq:indicator-rowsum})-(\ref{eq:indicator-decreasing})
are not necessary to solve the optimization problem, but are added to reduce
the feasible solution search space.

Practically, when solving (\ref{eq:objective}) on a commercial solver, we also
introduce an $\ell_2$ regularization term for $\varsegmentprice$ and
$\varsegmentpower$ to the objective to improve numerical stability.

\section{Results}

\subsection{Curve Interpretation for Multivariate Gaussian Scenarios}

In this section, we show the results of running the MIQP described in Section
\ref{sec:miqp} on two different synthetic Gaussian distributions of scenarios
to examine the behavior of the program and demonstrate a qualitative method
of interpreting the curve shapes. Table
\ref{table:simulation-parameters-synthetic} describes the simulation parameters
for the multivariate Gaussian used to generate the two cases. We denote the
case where wind and real-time prices are negatively correlated as Case 1,
and the case where they are positively correlated as Case 2. For both curves,
we set the program to generate six offer curve segments and then remove
segments that are smaller than 1 MW in width to improve readability.

\begin{figure*}[t!]
  \centering
\begin{subfigure}[t]{\columnwidth}
  \centering
  \includegraphics[width=0.9\columnwidth]{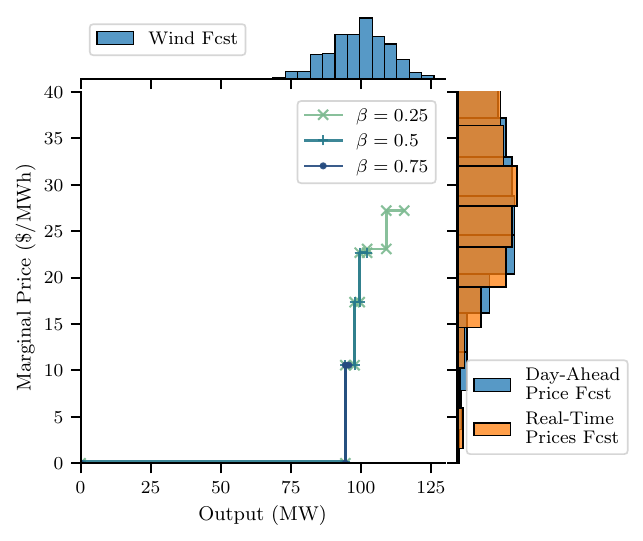}
  \caption{Offer curve for Case 1.}
    \label{fig:offer-curve-1}
\end{subfigure}
~
\begin{subfigure}[t]{\columnwidth}
  \centering
  \includegraphics[width=0.9\columnwidth]{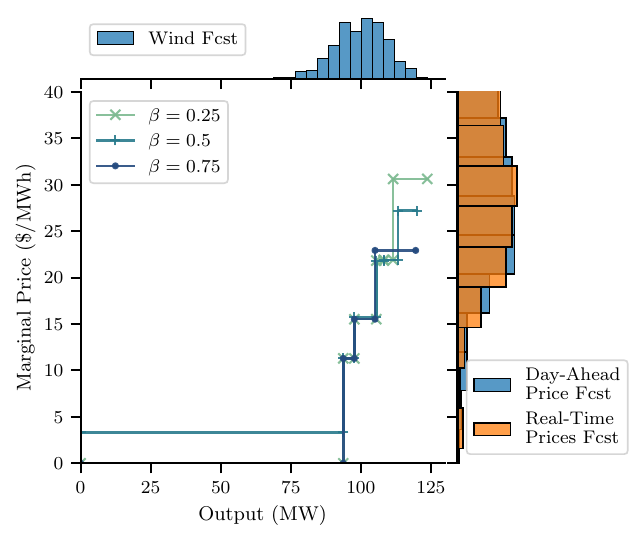}
  \caption{Offer curve for Case 2.}
    \label{fig:offer-curve-2}
\end{subfigure}

\begin{subfigure}[b]{\columnwidth}
  \includegraphics[width=0.9\columnwidth]{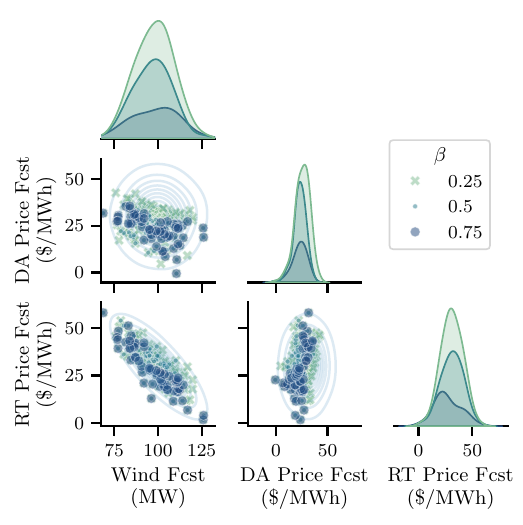}
  \caption{Active samples for Case 1 optimal solution.}
    \label{fig:jointplot-1}
\end{subfigure}
~
\begin{subfigure}[b]{\columnwidth}
  \includegraphics[width=0.9\columnwidth]{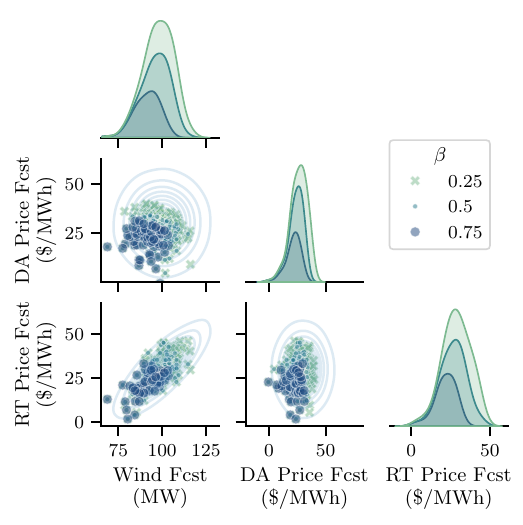}
  \caption{Active samples for Case 2 optimal solution.}
    \label{fig:jointplot-2}
\end{subfigure}
\caption{Offer curves and pairwise visualizations of distributions.}
\label{fig:curves-jointplots}
\end{figure*}

Fig. \ref{fig:curves-jointplots} shows the offer curves created by running the
MIQP for Cases 1 and 2. We observe that in both cases, the right side of the
offer curve reflects the nonzero variable costs associated with buyback
uncertainty, as would be expected.

For Case 1, increasing $\parambeta$ (more risk averse) reduces
the amount of total power offered in the day-ahead; the total amount offered
is larger than would come from just truncating the marginal wind distribution
at the corresponding $\parambeta$ alone, since there are scenarios where a
power shortfall may not result in significant financial loss if the real-time
price is also not high.

For Case 2, increasing $\parambeta$ decreases the price offered at the
right side of the curve when compared to a lower risk aversity. This result
is initially counterintuitive; one would generally expect that a higher
risk aversity would result in the uncertain wind portion in the 100-125MW
range to be priced higher, since that portion will be more likely to be
subject to buyback. However, in Case 2, real-time prices and wind are
positively correlated, meaning that any wind shortfall will actually be
repurchased a lower price in real-time than was generated through clearing
at a higher price in the day-ahead market; put differently, in expectation
there is no financial penalty for underdelivery, and the bulk of the loss
which is targeted by optimizing CVaR is due to insufficient revenue clearing in
the day-ahead market. We can qualitatively justify this claim by examining the
joint distribution plot in Fig. \ref{fig:jointplot-2}, which shows the
``active'' samples of day-ahead price, real-time price, and wind that
constitute the tail of the profit distribution that CVaR is calculated from.
At the higher $\parambeta$ values, the active samples are clustered where
day-ahead prices, real-time prices, and wind outputs are all lower than the
mean. Since the positive covariance means that there is less penalty for
underdelivery, the offer curve is constructed to clear power that may
not be deliverable in the real-time market to gain the additional day-ahead
revenue.

\renewcommand*{\arraystretch}{1.1}
\begin{table}[t!]
  \caption{Simulation parameters for synthetic Gaussian data cases}
  \label{table:simulation-parameters-synthetic}
  \begin{center}
  % \begin{adjustbox}{width=\columnwidth}
  \begin{tabular}{
    p{0.2\textwidth}>{\centering}p{0.1\textwidth}>{\centering\arraybackslash}p{0.1\textwidth}
  }
  \toprule \\
  Parameter & Symbol & Value \\
  \midrule \\
  Mean day-ahead price & $\mu_{DA}$ & 30 \$/MWh \\
  Mean real-time price & $\mu_{RT}$ & 30 \$/MWh \\
  Mean wind forecast & $\mu_{wind}$ & 100 MW \\
  Variance, day-ahead price & $\sigma^2_{DA}$ & 100 \\
  Variance, real-time price & $\sigma^2_{RT}$ & 100 \\
  Variance, wind forecast & $\sigma^2_{wind}$ & 100  \\
  Covariance*, RT price \& wind & $cov_{RT, wind}$ & $\pm$ 80 \\
  Number of scenarios generatred & $N_{\setscenario}$ & 250 \\
  \bottomrule \\
  \multicolumn{3}{l}{\scriptsize *All other covariances are zero.}
  \end{tabular}
  \end{center}
\end{table}

\section{NYISO Case Study}
We demonstrate the practicality of our program on real-world price data
from the New York Independent System Operator and synthetic wind forecast data
corresponding to a wind farm in the NYC zone. We compare the profits generated
when the MIQP is run for varying $\parambeta$ against a naive offer strategy
that offers at zero marginal price in the day-ahead market, up to a percentile
quantity of the forecast wind powers $\rvwind$. Hourly-level simulations are run
for the period of October 1st -- 31st, 2019.

\subsection{Scenario Generation}
For each operating day, we generate price samples by taking the historical
day-ahead and real-time prices in the corresponding hour over the
preceeding 50 calendar days. These price scenarios are paired with 100 wind
forecast scenarios generated from historical forecast data using PGscen
\cite{pgscen} to create 100 total scenarios for the hour of interest.

\subsection{Results}
For both the ``MIQP CVaR'' offer strategy and the ``forecast percentile''
strategy, the offer strategy is used to create the day-ahead offer curve. After
the day-ahead clearing quantity is determined from the actual day-ahead price,
the shortfall (or excess) wind production relative to the day-ahead clearing
quantity is bought-back (sold) at the real-time price.

To account for daily variations in revenue, we subtract the total profit from
the day-ahead and real-time markets under these two strategies from an
idealized profit number calculated by selling all the realized wind production
in whichever market is more profitable in hindsight. We refer to this
difference as the regret. For example, if the actual day-ahead hourly price was
\$10/MWh, the actual real-time price was \$20/MWh, and the actual maximum
wind production was 100 MW for that hour, the ideal profit would be \$200 for
that hour. If in that hour the MIQP CVaR strategy yielded a total profit of
\$150, the regret would be $\$50 = \$200 - \$150$.

\begin{figure}[t!]
  \centering
  \includegraphics[width=\columnwidth]{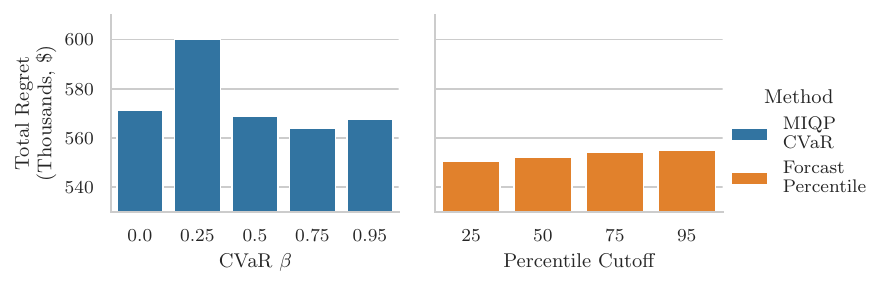}
  \caption{Strategy performance considering the sum of regret.}
    \label{fig:nyiso-regret-sum}
\end{figure}

Fig. \ref{fig:nyiso-regret-sum} shows the total regret over different choices
of risk preference $\parambeta$. There is no directional influence of
$\parambeta$ on the total regret. The MIQP CVaR approach also underperforms
the percentile-based offer strategy in the day-ahead market.

\begin{figure}[t!]
  \includegraphics[width=\columnwidth]{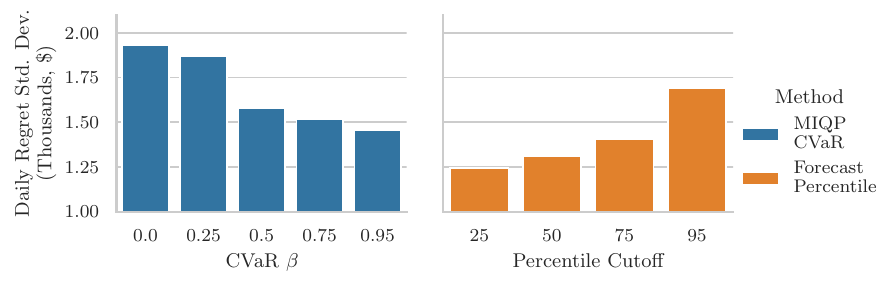}
  \caption{Strategy performance considering the standard deviation of regret.}
    \label{fig:nyiso-regret-std}
\end{figure}

Fig. \ref{fig:nyiso-regret-std} shows the average of the standard
deviations of the daily regret. Increasing the $\parambeta$ value (more risk
averse) decreases the variance of the daily regret, as expected. However, the
MIQP CVaR method still underperforms the naive 25\% and 50\% forecast percentile
methods in both total regret and variance of regret, indicating that there
is still room for improvement in the MIQP CVaR approach, possibly through
improve scenario creation.

\section{Computation \& Software}

All computations were performed on a laptop with an Intel Core i5-8300H CPU
(2.30GHz) and 31.2 GB of memory. The model was formulated using JuMP v1.15.1
in Julia v1.9.2. Gurobi v10.0.3 was used to solve the MIQP. Table
\ref{table:simulation-timing} describes the time to solve the model for a range
of scenario sample sizes; this time excludes model creation time. All models
were solved to optimality under the default Gurobi MIP gap of $10^{-4}$.

\section{Concluding Remarks}

In this work, we present an MIQP for valuing the CVaR-calculated risk of
different portions of a wind generator's offer curve, along with a pairwise
distribution plot to interpret the shapes of individual offer curves by
examining the set of scenarios that influence the calculation of those curves.
We introduce additional constraints to reduce the integer search space to
improve computation time. Empirical results using synthetic wind forecasts and
real-world NYISO price data demonstrate that adjusting the $\parambeta$ factor
of the MIQP CVaR strategy can adjust the variances in daily profits with little
tradeoff in total profits, although for the example dataset, such a strategy
underperformed a naive forecast percentile offer method in both total
profits and reduced variance of those profits. Such shortcomings could be
because there is insufficient price volatility in today's electricity markets
to reward hedging uncertainties. Further work should examine the price forecast
improvements which can be made to improve the quality of the CVaR-calculated
offer curve in reducing both the total and variance of regret.

\renewcommand*{\arraystretch}{1.1}
\begin{table}[t!]
  \caption{MIQP simulation times, Six curve segments}
  \label{table:simulation-timing}
  \begin{center}
  \begin{tabular}{
    cc
  }
  \toprule \\
  Number of scenarios & Median solve time (s)\\
  \midrule \\
  50  &  6.8 \\
  100  &  22.7 \\
  250  &  143.2 \\
  500  &  720.3 \\
  \bottomrule \\
  \end{tabular}
  \end{center}
\end{table}

\bibliographystyle{IEEEtran}
\bibliography{IEEEabrv,references/pes-gm-bibliography}

% Generated by IEEEtran.bst, version: 1.14 (2015/08/26)
\begin{thebibliography}{1}
\providecommand{\url}[1]{#1}
\csname url@samestyle\endcsname
\providecommand{\newblock}{\relax}
\providecommand{\bibinfo}[2]{#2}
\providecommand{\BIBentrySTDinterwordspacing}{\spaceskip=0pt\relax}
\providecommand{\BIBentryALTinterwordstretchfactor}{4}
\providecommand{\BIBentryALTinterwordspacing}{\spaceskip=\fontdimen2\font plus
\BIBentryALTinterwordstretchfactor\fontdimen3\font minus \fontdimen4\font\relax}
\providecommand{\BIBforeignlanguage}[2]{{%
\expandafter\ifx\csname l@#1\endcsname\relax
\typeout{** WARNING: IEEEtran.bst: No hyphenation pattern has been}%
\typeout{** loaded for the language `#1'. Using the pattern for}%
\typeout{** the default language instead.}%
\else
\language=\csname l@#1\endcsname
\fi
#2}}
\providecommand{\BIBdecl}{\relax}
\BIBdecl

\bibitem{curves-wind-bids}
A.~A. Thatte, L.~Xie, D.~E. Viassolo, and S.~Singh, ``Risk {{Measure Based Robust Bidding Strategy}} for {{Arbitrage Using}} a {{Wind Farm}} and {{Energy Storage}},'' \emph{IEEE Transactions on Smart Grid}, vol.~4, no.~4, pp. 2191--2199, Dec. 2013.

\bibitem{curves-demand}
G.~Ruan, H.~Zhong, B.~Shan, and X.~Tan, ``Constructing {{Demand-Side Bidding Curves Based}} on a {{Decoupled Full-Cycle Process}},'' \emph{IEEE Transactions on Smart Grid}, vol.~12, no.~1, pp. 502--511, Jan. 2021.

\bibitem{curves-price-taker-norway}
S.-E. Fleten and E.~Pettersen, ``Constructing bidding curves for a price-taking retailer in the norwegian electricity market,'' \emph{IEEE Transactions on Power Systems}, vol.~20, no.~2, pp. 701--708, May 2005.

\bibitem{yin-risk-based-dispatch}
X.~Yin, M.~D. Ili{\'c}, and B.~Sinopoli, ``Toward design of risk-based real-time dispatch at value,'' in \emph{2015 {{IEEE Power}} \& {{Energy Society Innovative Smart Grid Technologies Conference}} ({{ISGT}})}, Feb. 2015, pp. 1--5.

\bibitem{shen-acc}
D.~Shen and M.~Ilic, ``Valuing {{Uncertainties}} in {{Wind Generation}}: {{An Agent-Based Optimization Approach}},'' in \emph{2023 {{American Control Conference}} ({{ACC}})}, May 2023, pp. 1237--1242.

\bibitem{wind-power-offer-strategy-shin}
H.~Shin, D.~Lee, and R.~Baldick, ``An {{Offer Strategy}} for {{Wind Power Producers That Considers}} the {{Correlation Between Wind Power}} and {{Real-Time Electricity Prices}},'' \emph{IEEE Transactions on Sustainable Energy}, vol.~9, no.~2, pp. 695--706, Apr. 2018.

\bibitem{market-manipulation}
``The {{Guide}} to {{Energy Market Manipulation}},'' {Global Competition Review}, Tech. Rep., 2018.

\bibitem{cvar-paper}
R.~T. Rockafellar and S.~Uryasev, ``Optimization of conditional value-at-risk,'' \emph{The Journal of Risk}, vol.~2, no.~3, pp. 21--41, 2000.

\bibitem{pgscen}
R.~Carmona and X.~Yang, ``Joint {{Stochastic Model}} for {{Electric Load}}, {{Solar}} and {{Wind Power}} at {{Asset Level}} and {{Monte Carlo Scenario Generation}},'' Sep. 2022.

\end{thebibliography}
\end{document}